\documentclass[aps,showpacs,twocolumn,superscriptaddress]{revtex4}

\usepackage{graphicx}
\usepackage{epsfig}
\usepackage{amsfonts}
\usepackage{amsmath,amssymb}

\newcommand{\be}{\begin{equation}}
\newcommand{\ee}{\end{equation}}
\newcommand{\bn}{\begin{eqnarray}}
\newcommand{\en}{\end{eqnarray}}
\newcommand{\ba}{\begin{array}}
\newcommand{\ea}{\end{array}}
\newcommand{\bc}{\begin{center}}
\newcommand{\ec}{\end{center}}
\newcommand{\bml}{\begin{mathletters}}
\newcommand{\eml}{\end{mathletters}}

\begin{document}

\setcounter{topnumber}{4}
\setcounter{bottomnumber}{4}
\setcounter{totalnumber}{10}

\title{Surface-peaked effective mass in the nuclear energy density functional
and its influence on single-particle spectra}

\author{M. Zalewski}
\affiliation{Institute of Theoretical Physics, University of Warsaw, ul. Ho\.za 69, 00-681 Warsaw, Poland.}

\author{P. Olbratowski}
\affiliation{Institute of Theoretical Physics, University of Warsaw, ul. Ho\.za 69, 00-681 Warsaw, Poland.}

\author{W. Satu{\l}a}
\affiliation{Institute of Theoretical Physics, University of Warsaw, ul. Ho\.za 69, 00-681 Warsaw, Poland.}

\date{\today}

\begin{abstract}
Calculations for infinite nuclear matter with realistic nucleon-nucleon interactions suggest
that the isoscalar effective mass of a nucleon at the saturation density, $m^*/m$, equals $0.8\pm 0.1$.
This result is at variance with empirical data on the level density in finite nuclei, which are
consistent with $m^*/m\approx 1$. Ma and Wambach suggested that these two contradicting results may be
reconciled within a single theoretical framework by assuming a radial-dependent effective mass,
peaked at the nuclear surface. The aim of this exploratory work is to investigate this idea within the
density functional theory by using a Skyrme-type local functional enriched with new terms,
$\tau (\mathbf{\nabla}\rho)^2$ and $\tau\frac{d\rho}{dr}$, where $\tau$ and $\rho$ denote the kinetic
and particle densities, respectively. We show that each of these terms can give rise to a surface peak
in the effective mass, but of a limited height. We investigate the influence of the radial profile of
the effective mass on the spin-orbit splittings and centroids. In particular, we
demonstrate that the $\tau \frac{d\rho}{dr}$ term quenches the $1f_{5/2}-1f_{7/2}$ splitting in $^{40}$Ca,
which is strongly overestimated within conventional Skyrme parametrizations.
\end{abstract}


\maketitle

\section{Introduction}

Irrespective of their level of sophistication, infinite nuclear matter (INM) calculations provide a relatively
low isoscalar effective mass (IEM), $m^*/m \approx 0.8\pm 0.1$, at the saturation
density~\cite{[Bru58],[Jeu76w],[Fri81],[Wir88],[Zuo99]}. This finding is at variance with
empirical data on the level density at the Fermi energy in finite nuclei, which are consistent
with $m^*/m \approx 1$. Indeed, the level density close to the Fermi surface appears to be very well
described within simple models using bare nucleonic mass and a phenomenological one-body potential,
$U(\mathbf{r})$, of, e.g., the Woods-Saxon form~\cite{[Bro63],[Ber68],[Bar81],[Shl92],[Mug98]}.

It is not at all obvious, however, whether the success of such non-self-consistent approach is
built upon solid theoretical foundations or is a mere effect of the fitting strategy of $U(\mathbf{r})$,
which is adjusted precisely to the single-particle (SP) spectra. Indeed, rigorous and in principle exact
treatment of the single-particle propagator (Green's function) is obtained within the Landau quasi-particle
theory by solving the Dyson equation, which contains, in general, a spatially non-local and energy (frequency)
dependent self-energy~\cite{[Mig67],[Dic05]}. Hence, from theory point of view, the use of static,
momentum and energy (frequency) independent mean potentials seems to have no justification at all.

Quantitatively, the most important effect  on the SP spectra is expected to come from spacial
non-locality or momentum dependence of the mean potential. This effect is naturally taken into
account in the Hartree-Fock (HF) approximation based on effective nucleon-nucleon (NN) interactions,
like the most popular Skyrme \cite{[Sky56xw]} or Gogny \cite{[Gog75aw]} forces. Indeed, the zero-range
Skyrme (pseudo-) potential depends explicitly on momentum. In case of the finite-range Gogny interaction,
the effect enters through the exchange term. The energy dependence is considered only as a perturbation
correction atop of the HF
result~\cite{[Rin73],[Ham76aw],[Ber80w],[Wam82],[Som82]} and is usually completely ignored.

The self-consistent HF method involving the Skyrme or Gogny interactions leads to a non-uniform,
position-dependent IEM, $m^*({\mathbf r})/m$. Due to a rather restricted form of these forces, a
typical IEM profile changes smoothly from the bare-mass value of $1$ outside the nucleus to the
INM value of $0.8\pm 0.1$ in the nuclear interior. This lowers the nuclear level density at the
Fermi surface usually well below the experimental value and, in turn, spoils the spectroscopic
properties of these forces, see~\cite{[Rin80],[Sat08]} and references quoted therein.

The  contradicting results concerning the SP level density in INM and finite nuclei are
related to the fact that in finite nuclei the SP levels couple to low-lying collective
surface vibrations, which leads to an increase of the IEM at the surface. Several authors
argued that this effect can be taken into account within the random phase approximation,
i.e., by going beyond mean field, see~\cite{[Ham76],[Ber80w],[Lit06w]}. This viewpoint
is difficult to reconcile with the density functional theory (DFT) that should warrant
a proper radial dependence of the IEM through the fit to empirical data provided that the energy
density functional (EDF), or the effective NN interaction, is rich enough to accommodate the
anticipated increase of the IEM in the surface area.

Taking into account the particle-vibration coupling may help resolve the discussed controversy
in the following way. The IEM is smaller than one inside the nucleus, as required for nuclear matter.
On the other hand, it is peaked up at the surface, so that it may give one when averaged over the
nuclear volume. This may allow to simultaneously reproduce the level density in finite nuclei.

The surface-peaked IEM was first explored by Ma and Wambach. In their seminal paper~\cite{[Ma83]},
they used a phenomenological one-body Hamiltonian,
\begin{equation}
\label{MFHwEM}
H(\mathbf{r})=-\mathbf{\nabla}\cdot\frac{\hbar^2}{2m^*(\mathbf{r})}\mathbf{\nabla}+{U}(\mathbf{r}),
\end{equation}
consisting of the Woods-Saxon potential with the spin-orbit (SO) term, ${U}(\mathbf{r})$, and
the kinetic term with a radius-dependent IEM taken as follows,
\begin{equation}
\label{Ma_em}
\frac{m^*(\mathbf{r})}{m}=\left(1-\alpha g(\mathbf{r})\right)\left(1 + \beta \frac{dg(\mathbf{r})}{dr}\right),
\end{equation}
and depending on two adjustable constants, $\alpha$ and $\beta$, respectively. The function
$g(\mathbf{r})$ is related to unperturbed self-energy,
\begin{equation}
\label{Ma_g}
g(\mathbf{r})=\Sigma_0(\mathbf{r},\mathbf{k_F}(\mathbf{r}),\epsilon_F)/\Sigma_0(0,\mathbf{k_F}(0),\epsilon_F).
\end{equation}
It mimics the radial dependence of the SP potential and was chosen so that
$\Sigma_0(\mathbf{r},\mathbf{k_F}(\mathbf{r}),\epsilon_F) = U (\mathbf{r}) $. It essentially
follows the profile of the nucleonic density, $g(\mathbf{r}) \sim \rho (\mathbf{r})$. Ma and Wambach
demonstrated that one can indeed unify the description of the SP levels in INM and finite nuclei
by taking $m^*({\mathbf r})/m \approx 0.7$ in the bulk and peaking it up at the nuclear surface to
$m^*({\mathbf r})/m \approx 1.2 \div 1.5$, depending on the nucleus. This result was later confirmed by
Farine~\textit{et al.}~\cite{[Far01]} to hold in self-consistent HF calculations.

The aim of the present work is to explore the IEM modifications caused by the spacial non-locality
of the NN interaction and by the particle-vibration coupling within a fully self-consistent model.
Unlike Farine~\textit{et al.}~\cite{[Far01]}, who obtained a surface-peaked IEM by adding new terms
to the Skyrme interaction, we approach the problem from the perspective of the DFT, taking as a starting
point the local Skyrme functional and adding new terms thereto. Nowadays, the nuclear EDFs are
treated as basic entities, independent of the NN interaction. In this sense, our approach is
more general and more flexible than the Skyrme HF theory.

In this exploratory work, we extend the Skyrme functional by adding two terms, which
are described in detail in Sect.~\ref{terms}. Their effect on the radial IEM profile
is presented in Sect.~\ref{dziubek}. In Sect.~\ref{spectra}, we examine their influence
on the SP levels. Sect.~\ref{discussion} contains a discussion of the results, particularly
in connection with the recent studies of the tensor term. The paper is concluded in Sect.~\ref{summary}.
All considerations presented in this work concern the isoscalar effective mass, and are limited to
doubly-magic $N=Z$ nuclei. Such nuclei are spherical, which further simplifies the calculations.

\section{Extended Local Energy Density Functional}\label{terms}

The local Skyrme EDF, ${\mathcal H}({\mathbf r})$,
consists of the kinetic and potential parts,
\begin{equation}
\label{eq108}
{\mathcal H}({\mathbf r})=\frac{\hbar^2}{2m}\tau_0+\sum_{t=0,1} {\mathcal H}_t({\mathbf r}) .
\end{equation}
For even-even nuclei,
\begin{eqnarray}
\label{hte}
\mathcal{H}_t & = & C^{\rho}_t[\rho_0] \rho^2_t + C^{\Delta \rho}_t\rho_t\Delta\rho_t + \\ \nonumber
&\quad & C^{\tau}_t\rho_t\tau_t+C^J_t {\mathbb J}^2_t+C^{\nabla J}_t \rho_t {\mathbf \nabla}\cdot{\mathbf  J}_t .
\end{eqnarray}
Index $t=0,1$ denotes isospin, and $C_t$ are adjustable coupling constants. The constant $C^{\rho}_t[\rho_0]$
is density-dependent and equals
$C^{\rho}_t[\rho_0]=C^{\rho\,\prime}_t+C^{\rho\,\prime\prime}_t
(\frac{\rho_0}{\rho_{0,eq}})^{\alpha}$,
where $\rho_{0,eq}$ is the INM saturation density. The potential energy terms are bilinear
forms of time-even particle, kinetic, and tensor densities, $\rho_t$, $\tau_t$, ${\mathbb J}_t$,
and their derivatives. The density ${\mathbf J}_{t}$ is the vector part of the spin-current
tensor, ${\mathbf J}_{t,\lambda}=\sum_{\mu\nu}\epsilon_{\lambda\mu\nu}{\mathbb J}_{t,\mu\nu}$.
Since we consider $N=Z$ nuclei, only the isoscalar ($t=0$) channel is active. Readers interested
in details are referred, e.g., to Ref.~\cite{[Ben03]}.

Within this formalism, the IEM comes from variation of ${\mathcal H}$ over $\tau_0$,
\begin{equation}
\label{m-sky}
\frac{\hbar^2}{2m^*({\mathbf r})}=\frac{\delta{\mathcal H}}{ \delta\tau_0}=\frac{\hbar^2}{2m}+C^{\tau}_0\rho_0({\mathbf r}) .
\end{equation}
Thus, $m^*({\mathbf r})/m$ depends on the particle density. It is equal to one outside the nucleus and,
for positive values of $C_0^{\tau}$, smaller than one in the bulk. Therefore, the conventional Skyrme EDF
contains only the volume term $\sim g({\mathbf r})$ of parametrization (\ref{Ma_em}).

In order to allow for a surface-peaked profile of the IEM one has to add new terms to the functional.
Having an almost complete liberty in choosing their form, we decided to add either the term
\begin{equation}
\label{tgr}
{\mathcal H}_0^{(A)}({\mathbf r})=C_0^{\tau (\mathbf{\nabla}\rho)^2}\tau_0 ({\mathbf r})\left( \mathbf{\nabla}\rho_0({\mathbf r})\right)^2
\end{equation}
or
\begin{equation}
\label{tdr}
{\mathcal H}_0^{(B)}({\mathbf r})=C_0^{\tau d\rho/dr}\tau_0({\mathbf r})\frac{d\rho_0({\mathbf r})}{dr} .
\end{equation}
We call these two variants of the model A and B, and explore them separately, never including both terms simultaneously.

The rationale behind these two choices is almost self-evident. Both terms are proportional to the kinetic density,
$\tau_0$, and to derivatives of the particle density, $\rho_0$. Hence, their variations give
\begin{equation}
\label{m-tgr}
\frac{\hbar^2}{2m^*({\mathbf r})}=\frac{\hbar^2}{2m}+C^{\tau}_0\rho_0({\mathbf r})+C_0^{\tau(\mathbf{\nabla}\rho)^2}(\mathbf{\nabla}\rho_0({\mathbf r}))^2
\end{equation}
and
\begin{equation}
\label{m-tdr}
\frac{\hbar^2}{2m^*({\mathbf r})}=\frac{\hbar^2}{2m}+C^{\tau}_0\rho_0({\mathbf r})+C_0^{\tau d\rho/dr}\frac{d\rho_0({\mathbf r}) }{dr} ,
\end{equation}
for variants A and B, respectively. They modify the Skyrme value (\ref{m-sky}) at the nuclear surface, where
the derivative of the particle density is large, and give small contributions elsewhere. Since the density gradient
is negative in the surface region, while its square is positive, the peak is obtained for negative values of
$C_0^{\tau(\mathbf{\nabla}\rho)^2}$ in variant A, and for positive values of $C_0^{\tau d\rho/dr}$ in variant B.

The new terms also contribute to the mean potential according to the formula
\begin{equation}
U_0({\mathbf r})=\frac{\delta{\mathcal H}}{\delta\rho_0}.
\end{equation}
This contribution is included self-consistently in our calculations.

The term A is a scalar, so its form is natural even if spherical symmetry is not assumed.
The use of term B is restricted to nuclei of spherical shapes, what seems to
be a disadvantage. We have decided to explore this
term mainly because it mimics closely bulk and surface terms
$\frac{m^*(\mathbf{r})}{m} \approx 1-\alpha g(\mathbf{r}) +
\beta \frac{dg(\mathbf{r})}{dr}$ of the Ma and Wambach
parametrization (\ref{Ma_em}).

The new terms introduced here are of completely different nature than those used by
Farine~\textit{et al.}~\cite{[Far01]}. In their model, the IEM is modified not
by the gradient of the density, but through its value,
\begin{equation}
\label{m-Farine}
\frac{\hbar^2}{2m^*({\mathbf r})}=\frac{\hbar^2}{2m}+C^{\tau}_0\rho_0({\mathbf r})+C_0^{\beta}\rho_0^{\beta+1}({\mathbf r}) ,
\end{equation}
where $\beta$ is a new parameter. In such an approach, the peak at the surface is not
due to the slope of the density, but because the density there equals approximately half of its value in the interior.

In a preliminary study~\cite{[Zal10]}, we also investigated two other terms, $\tau^2$ and $\tau\triangle\rho$.
These, however, do not provide the expected radial profile of the IEM, and are not considered here.

\section{Radial dependence of the isoscalar effective mass}
\label{dziubek}

We base all our calculations on the SkXc Skyrme parametrization of Ref.~\cite{[Bro98]}.
This force was designed with particular attention to describe the SP levels, and has
the IEM close to unity right from the beginning. We add our new terms to the SkXc functional,
and vary some of its coupling constants as described below. We perform the calculations for
three doubly-magic, $N=Z$, spherical nuclei, $^{40}$Ca, $^{56}$Ni, and $^{100}$Sn.

As already explained, we aim at obtaining an IEM profile which provides $m^*(\mathbf{r})/m$ lower than one inside the nucleus and peaked at the surface, so that its mean value is close to one. In order to fulfill the latter condition we require that
\begin{equation}
\label{m-unit}
 \int d^3 {\mathbf r}\,\frac{\rho_0({\mathbf r})}{A}\,\frac{m^*({\mathbf r})}{m}=1 ,
\end{equation}
where $A$ is the mass number. A similar constraint was used by Ma and Wambach. Note that while the new coupling constants, $C_0^{\tau(\nabla\rho)^2}$ and $C_0^{\tau d\rho/dr}$, are responsible for the peak, $C^{\tau}_0$ accounts for lowering the IEM inside the nucleus. Thus, the condition (\ref{m-unit}) can be satisfied by an appropriate balance between the new coupling constants and $C^{\tau}_0$.

We proceed with the calculations for each of the considered nuclei separately in the following way. In both variants, we scan a range of the appropriate new coupling constant, and, for each value thereof, we readjust $C^{\tau}_0$ to fulfill the constraint (\ref{m-unit}), keeping all the remaining coupling constants frozen at their SkXc values.

\begin{table}
\caption{Limiting values of the new coupling constants corresponding to the largest peak and anti-peak in $^{40}$Ca.}
\label{table}
\begin{center}
\begin{tabular}{c|c|c}
\hline\hline
                                    & Largest anti-peak & Largest peak \\
\hline
$C_0^{\tau(\nabla\rho)^2}$ & 300\,MeV\,fm$^7$ & -280\,MeV\,fm$^7$ \\
$C_0^{\tau d\rho/dr}$               & -30\,MeV\,fm$^6$ & 70\,MeV\,fm$^6$\\
\hline
\end{tabular}
\end{center}
\end{table}

\begin{figure}[t]
\begin{center}
\includegraphics[width=0.4\textwidth, clip]{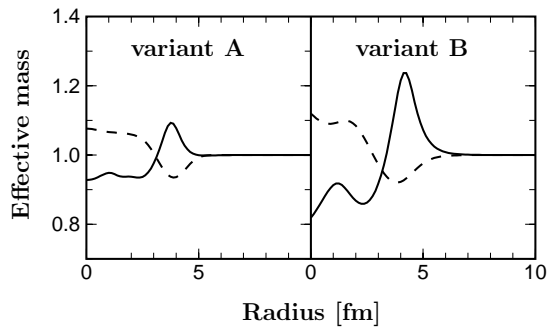}
\end{center}
\caption{Isoscalar effective mass, $m^*(r)/m$, versus $r$ in $^{40}$Ca for variants A (left) and B (right). The solid (dashed) lines show the IEM profiles with the largest peak (anti-peak), obtained for the limiting values of the new coupling constants shown in Table \ref{table}.}
\label{shapes}
\end{figure}

For completeness, we examine both positive and negative values of each new coupling constant, which corresponds both to a peak and to an anti-peak at the surface. It turns out that the range of each new constant is limited in both directions. Beyond that range, rapid oscillations in the density occur, and the HF iterations diverge. The limiting values of the coupling constants give rise to the largest peak or anti-peak that could be obtained. As an example, we give those limiting values for $^{40}$Ca in Table \ref{table}, and show the corresponding IEM profiles in Fig.~\ref{shapes}.

It can be seen from Fig.~\ref{shapes} that both variants of our model are rich enough to render a surface-peaked IEM. However, the height of the peak is rather limited. In variant A, the IEM in the nuclear interior is larger than $\sim$0.9 and reaches only $\sim$1.10 at the surface. These values equal $\sim$0.8 and $\sim$1.25 for variant B, respectively. For the same nucleus, Ma and Wambach quote the values of $\sim$0.7 and $\sim$1.25. This result can be reproduced only in variant B of our model. It should also be mentioned that we failed to obtain a surface-peaked IEM with those Skyrme parametrizations which have a small effective mass in the INM limit, like $\sim 0.7\div 0.8$.

\section{Influence of the isoscalar effective mass geometry on single-particle levels}
\label{spectra}

We are now in a position to examine the influence of the new terms on SP spectra.
Since our later discussion will mostly concern the SO properties, we consider the SO splittings,
\begin{equation}
E_{SO}(n,\ell)\equiv e(j=\ell-\tfrac{1}{2},n,\ell)-e(j=\ell+\tfrac{1}{2},n,\ell) ,
\end{equation}
and the centroids of the SO partners,
\begin{equation}
E_{C}(n,\ell)\equiv\frac{1}{2}\left[
e(j=\ell-\tfrac{1}{2},n,\ell)+e(j=\ell+\tfrac{1}{2},n,\ell)\right] .
\end{equation}
Here, $e(j,n,\ell)$ are the SP energies.
Since we are interested in delineating trends rather than details
concerning the influence of the new terms on SP spectra
we take $e(j,n,\ell)$ simply as eigen-energies of the one-body Hamiltonian,
see discussion in Ref.~\cite{[Zal08]}.
To focus attention, we discuss only the neutron splittings and centroids.

Let us remind that within the DFT formalism, the SO splittings are determined by the SO field,
$B_0({\mathbf r})$, which is a derivative of ${\mathcal H}$ over ${\mathbf
J}_{0}$, and, for spherical symmetry, equals
\begin{equation}
\label{so_field}
B_0(r)= C^J_0 J_0 - C^{\nabla J}_0\frac{d\rho_0}{dr} ,
\end{equation}
where $J_0$ stands for radial component of the vector part of the spin-current
${\mathbf J}_{0}$. Since in most Skyrme parametrizations, including SkXc, the
tensor coupling constant, $C^{\nabla J}_0$, is very small or zero, the SO
field is in practice proportional to the gradient of the density, which is
largest at the surface.

We perform two different types of calculations. In both cases we scan the whole admissible range of the new coupling constants, but:
\begin{enumerate}
\item For each value of the new coupling constant we readjust $C^{\tau}_0$ to fit the constraint (\ref{m-unit}),
as in Sect. \ref{dziubek}. Hereafter, we call these calculations {\it perturbative}.
\item In addition to adjusting $C^{\tau}_0$ to the condition (\ref{m-unit}), we attempt to refit
also other coupling constants to masses and radii of the considered nuclei. These calculations
will be dubbed {\it non-perturbative}, and are described in details below.
\end{enumerate}

\subsection{Perturbative calculations}
\label{pert}

\begin{figure}[t]
\begin{center}
\includegraphics[width=0.4\textwidth, clip]{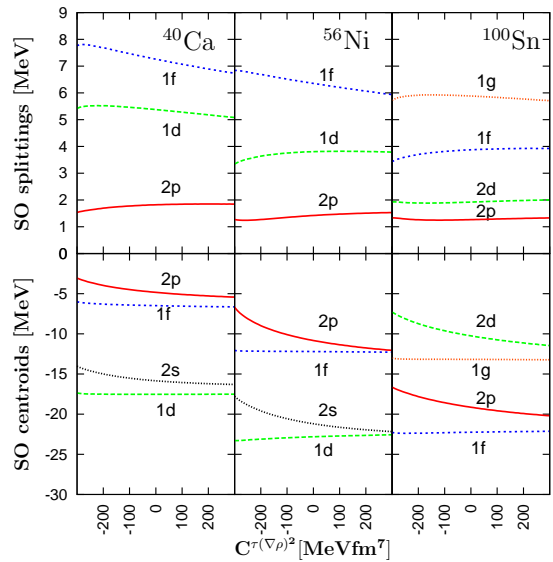}
\caption{(Color online) Spin-orbit splittings (upper panels) and centroids
(lower panels) in function of $C^{\tau(\nabla\rho)^2}$ obtained from variant A
of the {\it perturbative} calculations in $^{40}$Ca (left), $^{56}$Ni
(middle), and $^{100}$Sn (right).} \label{split_tgr1} \end{center}
\end{figure}

The SO splittings and centroids obtained from the {\it perturbative} calculations in variant A are
plotted in Fig.~\ref{split_tgr1} in function of $C_0^{\tau(\nabla\rho)^2}$. Note that the peak appears
for negative values of this coupling constant. As already discussed, the term A modifies the IEM
profile rather weakly. Consequently, it also affects the SO splittings and centroids in a relatively
modest way. Exceptions are the $1f_{5/2}-1f_{7/2}$ splittings in $^{40}$Ca and $^{56}$Ni,
and the $n=2$ centroids, $2s$, $2p$, and $2d$.

\begin{figure}[b]
\begin{center}
\includegraphics[width=0.4\textwidth,clip]{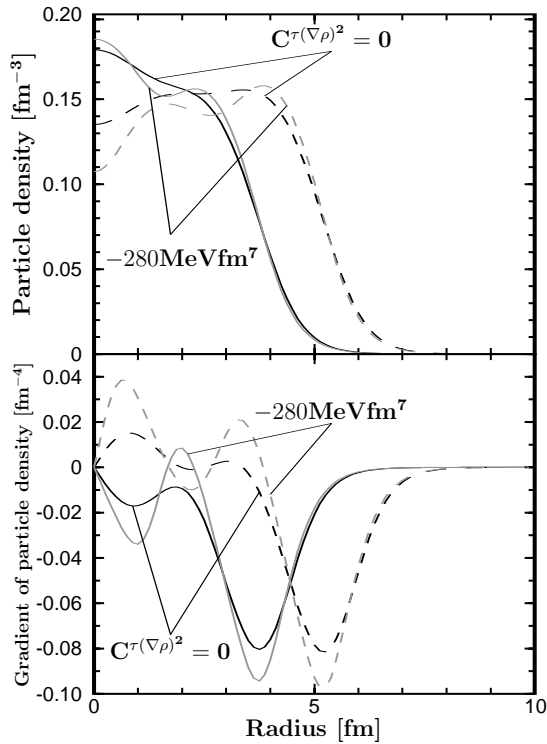}
\caption{Isoscalar particle density (upper panel) and its gradient (lower panel) versus radius obtained
from variant A of the {\it perturbative} calculations in $^{40}$Ca (solid lines) and $^{56}$Ni (dashed lines)
for $C^{\tau(\nabla \rho)^2}=0$ (black lines) and for the highest-peak value of
$C^{\tau(\nabla \rho)^2}=-280$\,MeV\,fm$^7$ (gray lines). }
\label{den_tgr}
\end{center}
\end{figure}

In order to understand these results, let us inspect Fig.~\ref{den_tgr}, which shows the particle
density, $\rho_0$, and its gradient in function of $r$ for $^{40}$Ca and $^{56}$Ni, calculated in variant A
for $C_0^{\tau(\nabla\rho)^2}$ equal to 0 and to the limiting value of -280\,MeV\,fm$^7$. It turns out that even the
largest possible peak only weakly affects the density profile in the surface region, but quite strongly amplifies the
density fluctuations inside the nucleus. Thus, the gradient of the density and the SO field are only moderately
increased at the surface, which does not influence the SO splittings significantly. On the other hand, the modifications
in the interior have more impact on low-$\ell$ states with wave functions inside the nucleus, than on higher-$\ell$ states,
whose wave functions are more spread toward the surface.

\begin{figure}[t]
\begin{center}
\includegraphics[width=0.4\textwidth, clip]{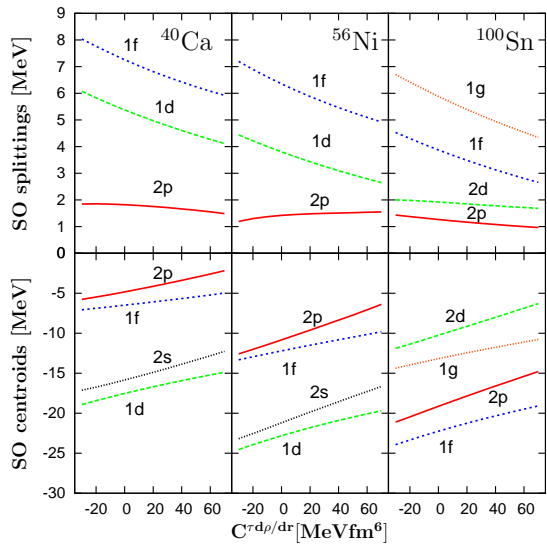}
\caption{(Color online) Spin-orbit splittings (upper panels) and centroids
(lower panels) in function of $C^{\tau d\rho/dr}$ obtained from variant B of
the {\it perturbative} calculations in $^{40}$Ca (left), $^{56}$Ni (middle)
and $^{100}$Sn (right).} \label{split_tdr1} \end{center} \end{figure}

The SO splittings and centroids obtained in variant B are presented in Fig.~\ref{split_tdr1}.
Note that here the peak appears for positive values of $C_0^{\tau d\rho/dr}$. The tendencies
are very different than in variant A. With the development of a surface-peaked IEM, one
observes a systematic and substantial reduction of the SO splittings for high-$\ell$ orbits.
In particular, the splittings for $n=1$ orbitals, like $1d$, $1f$, $1g$, change by as much as
$2$MeV throughout the entire range of $C_0^{\tau d\rho/dr}$. The splittings of the states with
$n=2$ and low$-\ell$, including $2s$, $2p$, $2d$, are relatively weekly affected by the new
term with no prevailing trend. However, these splittings are small, so their relative variations
can still be as large as $20$\%.
The centroids of all the calculated orbits go steadily up with $C_0^{\tau d\rho/dr}$.
There is also a tendency to increase the distance,
\begin{equation}
\Delta E=|E_C(n,\ell)-E_C(n+1,\ell-2)| ,
\end{equation}
between the SO centroids of, loosely speaking, the {\it pseudo-spin} partners.

\begin{figure}[t]
\begin{center}
\includegraphics[width=0.4\textwidth, clip]{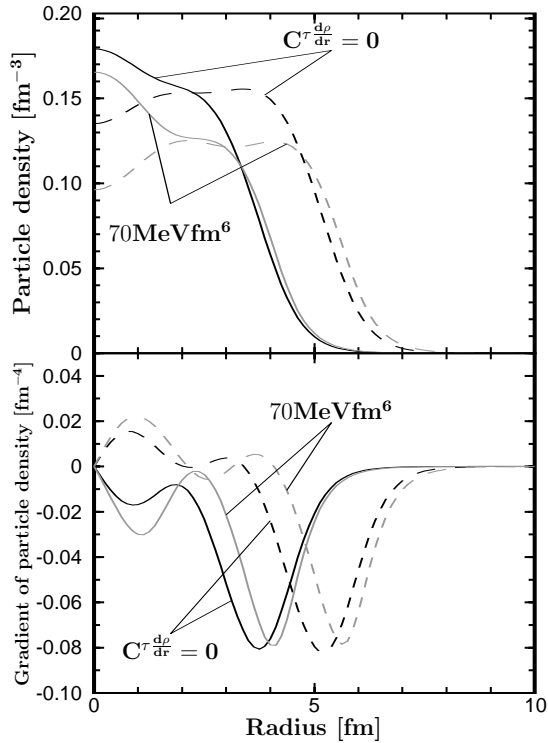}
\caption{Isoscalar particle density (upper panel) and its gradient (lower panel) versus radius obtained from variant B of the {\it perturbative} calculations in $^{40}$Ca (solid lines) and $^{56}$Ni (dashed lines) for $C^{\tau d\rho/dr}=0$ (black lines) and for the highest-peak value of $C^{\tau d\rho/dr}=70$\,MeV\,fm$^6$ (gray lines). }
\label{den_tdr}
\end{center}
\end{figure}

The changes of the SO splittings and centroids with $C_0^{\tau d\rho/dr}$ can be nicely correlated with the changes in the radial dependence of the particle density shown in Fig.~\ref{den_tdr}. With increasing $C_0^{\tau d\rho/dr}$, the density drops in the nuclear interior and the surface of the nucleus moves outward. Hence, also the gradient of the density and the SO field are repelled from the bulk, where the wave functions are located, and the SO splittings decrease. This argument is stronger for high$-\ell$, $n=1$ orbits with an appreciable amount of the wave function in the surface region, and much weaker for low$-\ell$, $n=2$ orbits, whose wave functions have one node inside the nucleus. The centroids move up in energy because the potential well becomes shallower and wider.

\subsection{Non-perturbative calculations}
\label{refit}

\begin{figure}[t]
\begin{center}
\includegraphics[width=0.35\textwidth, clip]{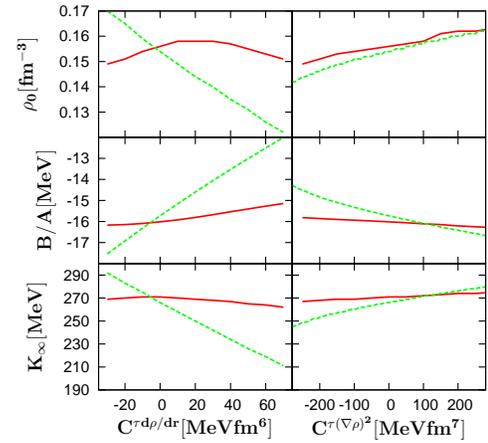}
\caption{(Color online) Saturation density (top), binding energy per nucleon (middle), and
incompressibility modulus (bottom) of infinite nuclear matter obtained from {\it perturbative} (green dashed lines) and {\it non-perturbative} (red solid lines) calculations in variant A (right) and B (left) in function of the corresponding new coupling constants.}
\label{nmp}
\end{center}
\end{figure}

In the {\it perturbative} calculations, the new terms deteriorate the performance of the SkXc functional concerning the INM saturation energy, binding energy per nucleon, and the incompressibility modulus, as well as radii, binding energies, and SP levels in finite nuclei. The influence on the INM properties is illustrated in Fig.~\ref{nmp}. This is so because the {\it perturbative} calculations take some coupling constants too far away from the SkXc values without readjusting the remaining ones. Hence, although helpful in understanding the response of the nucleus against the new terms, these calculations are of limited applicability. In this Section, we attempt to remove this drawback by refitting some other Skyrme coupling constants to selected radii, masses, and SP levels in finite nuclei.

We freeze the tensor and SO coupling constants, $C_0^J$ and $C_0^{\nabla J}$, at their SkXc values. At each value of $C_0^{\tau(\nabla\rho)^2}$ or $C_0^{\tau d\rho/dr}$, the constant $C_0^{\tau}$ is adjusted to fulfill the condition (\ref{m-unit}), as in the {\it perturbative} calculations. Simultaneously, $C_0^{\rho~\prime}$, $C_0^{\rho~\prime\prime}$, and $C_0^{\rho\Delta\rho}$, are refitted to minimize the merit function
\begin{equation}
\chi^2=\sum_i{\frac{(w_i^{(th)}-w_i^{(exp)})^2}{\sigma_i^2}} .
\end{equation}
Here, $w_i^{(th)}$ and $w_i^{(exp)}$ denote the theoretical and experimental values of the $i$-th observable, and the weight $\sigma_i^2$ is an {\it a priori} assumed error associated with the $i$-th observable. It may be understood as a desired theoretical accuracy. Observables used in the fit include masses and radii of $^{16}$O, $^{40}$Ca and $^{56}$Ni, and the mass of $^{100}$Sn. We take $\sigma_i^2$ equal to $2$MeV for masses and $0.1$fm for radii.

It is not our intention to create a new functional here. We only aim at investigating the impact of the new terms on the SP levels while keeping the basic nuclear properties close to the SkXc values. This is why we fix the coupling constants $C_0^J$ and $C_0^{\nabla J}$. Indeed, they influence the SO splittings and centroids directly through Eq. (\ref{so_field}). By fixing them, we can track the response of the splittings and centroids solely to the new terms.

In the {\it non-perturbative} calculations, the masses of $^{40}$Ca, $^{56}$Ni and $^{100}$Sn vary by less than $\pm$1\% throughout the whole range of the coupling constant $C_0^{\tau d\rho/dr}$ in variant B. The mass of $^{16}$O varies within $\pm$5\%.  The INM properties are shown in Fig.~\ref{nmp}. For most values of the new coupling constants, the saturation density stays within a range of $\pm 0.005fm^{-3}$ ($\pm$3\%) with respect to the SkXc value, the binding energy per particle changes by no more than $\pm 0.6$MeV ($\pm$4\%) and the incompressibility modulus by less than $\pm 5$MeV ($\pm$ 2\%). These results guarantee that our refitted functional retains the basic characteristics of the SkXc parametrization and that new effects, if any, can be ascribed mostly to the new terms. The IEM profiles obtained from the {\it non-perturbative} calculations do not differ significantly from those given in Fig.~\ref{shapes}.

\begin{figure}[t]
\begin{center}
\includegraphics[width=0.4\textwidth, clip]{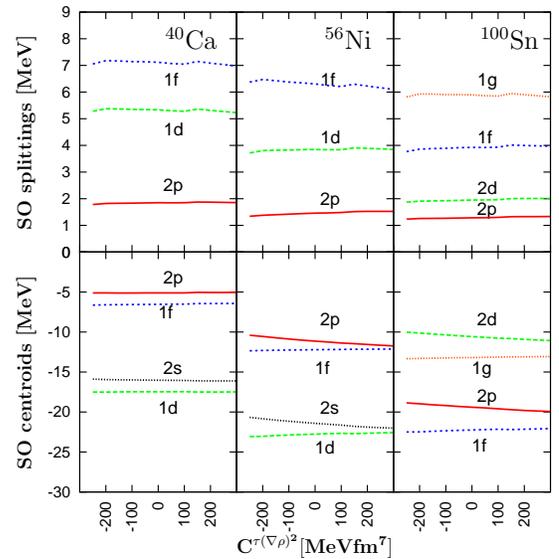}
\caption{(Color online) Spin-orbit splittings (upper panels) and centroids (lower panels) in function of $C^{\tau(\nabla\rho)^2}$ obtained from variant A of the {\it non-perturbative} calculations in $^{40}$Ca (left), $^{56}$Ni (middle), and $^{100}$Sn (right).}
\label{split_tgrf}
\end{center}
\end{figure}

\begin{figure}[t]
\begin{center}
\includegraphics[width=0.4\textwidth, clip]{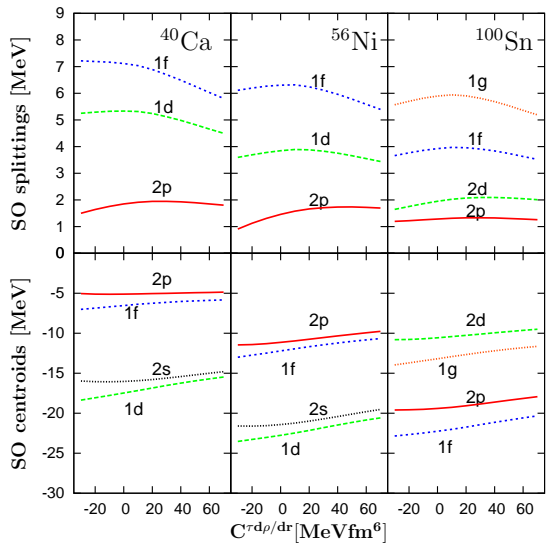}
\caption{(Color online) Spin-orbit splittings (upper panels) and centroids (lower panels) in function of $C^{\tau d\rho/dr}$ obtained from variant B of the {\it non-perturbative} calculations in $^{40}$Ca (left), $^{56}$Ni (middle) and $^{100}$Sn (right).}
\label{split_tdrf}
\end{center}
\end{figure}

The neutron SO splittings and centroids are shown in Figs.~\ref{split_tgrf} and~\ref{split_tdrf} for variants A and B, respectively. In variant A, there is almost no dependence of the splittings on $C_0^{\tau(\nabla\rho)^2}$, and only a small increase in $\Delta E$ in $^{56}$Ni and $^{100}$Sn occurring with the development of the peak. Variant B is more promising. Indeed, the previously observed trend of splittings decreasing with the onset of the peak is retained. The centroids move slightly up in energy and the values of $\Delta E$ are becoming systematically smaller.

\section{Discussion}
\label{discussion}

An important conclusion from our study concerns the behavior of the neutron $1f_{5/2}-1f_{7/2}$ splitting in $^{40}$Ca with the development of the surface peak in the IEM. This quantity is systematically overestimated in the conventional Skyrme functionals, independent of the parametrization, see Ref.~\cite{[Zal08]} and references therein. One can see, e.g., from Fig.~\ref{split_tdr1} for $C_0^{\tau d\rho/dr}$=$0$ that the SkXc functional misses the experimental value of $5.240$MeV~\cite{[Oro96w]} by more than $2$MeV. From this point of view, the results obtained in variant A of our model are rather disappointing because the concerned splitting increases even more with the onset of the peak in the {\it perturbative} calculations, which is definitely undesired, and remains almost constant in the {\it non-perturbative} calculations. Variant B is by far more promising. It can be seen from Fig.~\ref{split_tdr1} that for $C_0^{\tau d\rho/dr}$=$70$MeV\,fm$^6$, the difference between theory and experiment drops below $1$MeV in the {\it perturbative} calculations. This decreasing trend is retained also in the {\it non-perturbative} calculations.

These findings should be discussed in connection with the recent investigations of the tensor term in the Skyrme functional, see, e.g., Ref.~\cite{[Zal08],[Zal09],[Les07]}. The tensor term, $C^J_t{\mathbb J}_t^2$, together with the SO term, $C^{\nabla J}_t\rho_t{\mathbf\nabla}\cdot{\mathbf J}_t$, determine the SO field according to Eq.~(\ref{so_field}) and, consequently, the SO splittings. As explained, e.g., in Ref.~\cite{[Les07]}, the tensor term is inactive in spin-saturated nuclei, i.e., when the SO partners are both empty or filled, and is strongest in spin-unsaturated nuclei, i.e., when one SO partner is filled and one empty. From among the nuclei considered here, $^{40}$Ca is spin-saturated, while $^{56}$Ni and $^{100}$Sn are spin-unsaturated.

It was demonstrated in Refs.~\cite{[Zal08],[Zal09]} that in order to reproduce the experimental value of the neutron $1f_{5/2}-1f_{7/2}$ splitting in $^{40}$Ca, one has to dramatically reduce the SO strength, $C^{\nabla J}_t$. This, in turn, requires a substantial increase of the tensor strength, $C^J_t{\mathbb J}_t^2$, in order to keep the SO splittings in spin-unsaturated nuclei at a reasonable level. On the other hand, global fits to nuclear masses~\cite{[Les07]} do not allow for such significant modifications of the SO and tensor coupling constants.

From this point of view, the reduction of the neutron $1f_{5/2}-1f_{7/2}$ splitting in $^{40}$Ca by the surface-peaked IEM opens a possibility of reproducing this quantity without such a radical quenching of the SO strength. Consequently, also the tensor term would not need to be strengthened so much. This observations rise hopes that, with the inclusion of the surface-peaked IEM, the SP properties and nuclear masses may be both correctly described by a single functional.

Another interesting observation is that also the splittings between the {\it pseudo-spin} partners decrease with the emerging peak in variant B of the {\it non-perturbative} calculations. These values are also overestimated by conventional Skyrme functionals, so that their reduction may improve the agreement with experiment.

\section{Summary and conclusions}
\label{summary}

We considered the SkXc Skyrme functional enriched with two new terms, $C_0^{\tau(\nabla\rho)^2}\tau_0\left(\mathbf{\nabla}\rho_0\right)^2$ and $C_0^{\tau d\rho/dr}\tau_0\frac{d\rho_0}{dr}$, which we call A and B, respectively. They are designed to produce a surface-peaked radial profile of the isoscalar effective mass (IEM). Such a profile accounts for coupling of single-particle (SP) states to low-lying surface vibrations, and it may allow for a correct description of SP level density close to the Fermi surface both in infinite nuclear matter (INM) and in finite nuclei.

It was demonstrated that both terms do indeed produce a peak in the IEM, but of a limited height, because too large values of the new coupling constants lead to an instability and divergence of the HF iterations.

It turned out that the response of the SP levels to the presence of the peak strongly depends on the variant of the model. While the SP properties hardly change in variant A, most of the SO splittings systematically decrease with the development of the peak in variant B. This includes the neutron $1f_{5/2}-1f_{7/2}$ splitting in $^{40}$Ca, which is strongly overestimated by conventional Skyrme functionals. Thus, the inclusion of the new terms may help reproduce the empirical SO splittings. It may also be helpful in determining the correct strength of the tensor term, which is also partially responsible for the SO properties. Finally, the onset of the peak in variant B reduces the energy distance between the {\it pseudo-spin} partners in $^{40}$Ca, which also improves the agreement between theory and experiment.

Rigorous fits to bulk and SP data are necessary to corroborate or falsify our findings. However, this exploratory work demonstrates that construction of local functionals by adding specific, physically well motivated terms may constitute a promising alternative to the gradient-expansion method proposed recently in Ref.~\cite{[Car08]}.

\section{Acknowledgments}
\label{acknowledgements}

We would like to thank Janusz Skalski for inspiring discussions. This work was supported in part by the Polish Ministry of Science under Contracts No.~N~N202~239137 and~N~N202~328234.


\begin{thebibliography}{10}

\bibitem{[Bru58]}
{K.A. Brueckner and J.L. Gammel, Phys. Rev. {\bf 109}, 1023 (1982).}

\bibitem{[Jeu76w]}
{J.P. Jeukenne, A. Lejeunne, and C. Mahaux, Phys. Rep. {\bf 25}, 83 (1976).}

\bibitem{[Fri81]}
{B.A. Friedman and V.R. Pandharipande, Nucl. Phys. {\bf A361}, 502 (1981)}.

\bibitem{[Wir88]}
{R.B. Wiringa, V. Fiks, and A. Fabrocini, Phys. Rev. C {\bf 38}, 1010 (1988)}.

\bibitem{[Zuo99]}
{W. Zuo, I. Bombaci, U. Lombardo, Phys. Rev. C \textbf{60}, 024605 (1999)}.

\bibitem{[Bro63]}
{G.E. Brown, J.H. Gunn, and P. Gould, Nucl. Phys. {\bf 46}, (1963) 598}.

\bibitem{[Ber68]}
{T. Berggren, Nucl. Phys. {\bf A109}, 265 (1968)}.

\bibitem{[Bar81]}
{M. Barranco and J. Treiner, Nucl. Phys. {\bf A351}, 269 (1981)}.

\bibitem{[Shl92]}
{S. Shlomo, Nucl. Phys. {\bf A539}, 17 (1992)}.

\bibitem{[Mug98]}
{S.F. Mughabghab and C. Dunford, Phys. Rev. Lett {\bf 81}, 269 (1998)}.

\bibitem{[Mig67]}
{A.B. Migdal, {\sl Theory of Finite Fermi Systems and Applications to Atomic
  Nuclei} (Interscience, New York, 1967)}.

\bibitem{[Dic05]}
{W.H. Dickhoff and D. Van Neck, {\sl Many-body Theory Exposed! Propagator
  Description of Quantum Mechanics in Many-Body Systems} (World Scientific
  Publishing Co. Pte. Ltd., Singapore, 2005)}.

\bibitem{[Sky56xw]}
{T.H.R. Skyrme, Phil. Mag. {\bf 1} (1956) 1043; Nucl. Phys. {\bf 9} (1959)
  615.}

\bibitem{[Gog75aw]}
{D. Gogny, Nucl. Phys. {\bf A237} (1975) 399.}

\bibitem{[Rin73]}
{P. Ring and E. Werner, Nucl. Phys. {\bf A211}, 198 (1973)}.

\bibitem{[Ham76aw]}
{I. Hamamoto and P. Siemens, Nucl. Phys. {\bf A269}, 199 (1976).}

\bibitem{[Ber80w]}
{V. Bernard and N. Van Giai, Nucl. Phys. {\bf A348}, 75 (1980).}

\bibitem{[Wam82]}
{J. Wambach, V. Mishra, and Li Chu-hsia, Nucl. Phys. {\bf A380}, 285 (1982)}.

\bibitem{[Som82]}
{H.M. Sommermann, T.T.S. Kuo, and K.F. Ratcliff, Phys. Lett. {\bf 112B}, 108
  (1982)}.

\bibitem{[Rin80]}
{P. Ring and P. Schuck, {\sl The Nuclear Many-Body Problem} (Springer-Verlag,
  Berlin, 1980)}.

\bibitem{[Sat08]}
{W. Satu{\l}a, R.A. Wyss, and M. Zalewski, Phys. Rev. C {\bf 78}, 011302(R)
  (2008)}.

\bibitem{[Ham76]}
{I. Hamamoto, Phys. Lett. {\bf 61B} (1976) 343}.

\bibitem{[Lit06w]}
{E. Litvinova and P. Ring, Phys. Rev. C {\bf 73}, 044328 (2006).}

\bibitem{[Ma83]}
{Z.Y. Ma and J. Wambach, Nucl. Phys. {\bf A402}, 275 (1983)}.

\bibitem{[Far01]}
{M. Farine, J.M. Pearson, and F. Tondeur, Nucl. Phys. {\bf A696}, 396
(2001)}.

\bibitem{[Ben03]}
{M. Bender, P.-H. Heenen, and P.-G. Reinhard, Rev. Mod. Phys. {\bf 75}, 121
  (2003)}.

\bibitem{[Zal10]}
{M. Zalewski, P. Olbratowski, and W. Satu{\l}a, to be published in Int. J. Mod.
  Phys. {\bf E}}.

\bibitem{[Bro98]}
{B.A. Brown, Phys. Rev. C {\bf 58}, 220 (1998)}.

\bibitem{[Zal08]}
{M. Zalewski, J. Dobaczewski, W. Satu{\l}a, and T.R. Werner, Phys. Rev. C {\bf
  77}, 024316 (2008)}.

\bibitem{[Oro96w]}
{A. Oros, Ph.D. thesis, University of K\"oln, 1996.}

\bibitem{[Zal09]}
{M. Zalewski, P. Olbratowski, M. Rafalski, W. Satula, T.R. Werner, and R.A.
  Wyss, Phys. Rev. C {\bf 80}, 064307 (2009)}.

\bibitem{[Les07]}
{T. Lesinski, M. Bender, K. Bennaceur, T. Duguet, and J. Meyer, Phys. Rev. C
  {\bf 76}, 014312 (2007)}.

\bibitem{[Car08]}
{B.G. Carlsson, J. Dobaczewski, and M. Kortelainen, Phys. Rev. C {\bf 78},
  044326 (2008)}.

\end{thebibliography}

\end{document}